\documentclass[conference]{IEEEtran}
\IEEEoverridecommandlockouts
\usepackage{amsmath,amssymb,amsfonts}
\usepackage{algorithmic}
\usepackage{graphicx}
\usepackage{textcomp}
\usepackage{xcolor}
\usepackage{etoolbox}
\usepackage{amsthm}
\usepackage{physics2}
\usepackage{dsfont}
\usepackage{mathtools}
\usepackage{layouts}
\usepackage[colorlinks=true,urlcolor=teal,
            linkcolor=teal,citecolor=teal]{hyperref}
\usepackage[capitalise]{cleveref}
\usepackage{booktabs}
\usepackage{siunitx}
\usepackage{xspace}
\usepackage{pgffor}
\usepackage[doipre={DOI:}]{uri}

\usepackage[backend=biber,style=ieee,maxnames=3,citetracker=true,citestyle=ieee-comp,defernumbers=true]{biblatex}

\newbibmacro*{title+link}[1]{%
  \iffieldundef{doi}
    {%
      \iffieldundef{url}
        {#1}%
        {\href{\thefield{url}}{\textcolor{teal}{#1}}}%
    }
    {\href{https://doi.org/\thefield{doi}}%
       {\textcolor{teal}{#1}}}%
}

\DeclareFieldFormat
  [online,article,inbook,incollection,inproceedings,patent,thesis,unpublished,misc]
  {title}{%
    \usebibmacro{title+link}{#1}%
  }

\renewbibmacro*{doi+eprint+url}{}
\renewbibmacro*{url+urldate}{}

\AtEveryBibitem{
	\clearfield{pages}
	\clearfield{isbn}
	\clearfield{eprint}
	\clearfield{issn}
	\clearfield{abstract}
	\clearfield{volume}
	\clearfield{number}
	\clearfield{month}
}

\renewbibmacro{in:}{}
\AtBeginBibliography{\footnotesize}

\providetoggle{showtodos}
\settoggle{showtodos}{false}

\newtheorem{definition}{Definition}

\usephysicsmodule{ab, ab.braket, ab.legacy}

\newcommand{\renyi}{Rényi\xspace}
\newcommand{\eg}{\emph{e.g.}\xspace}
\newcommand{\ie}{\emph{i.e.}\xspace}

\DeclareMathOperator{\ketzero}{\ket|0>}

\DeclareMathOperator{\UIQP}{U_{\mathrm{IQP}}}
\DeclareMathOperator{\diag}{\mathrm{diag}}
\DeclareMathOperator{\sre}{\mathrm{SRE}}
\DeclareMathOperator{\jsd}{D_{\mathrm{JS}}}
\DeclareMathOperator{\dkl}{D_{\mathrm{KL}}}

\DeclareMathOperator{\Haar}{\mathrm{Haar}}

\DeclareMathOperator{\ketpsi}{\ket|\psi>}
\DeclareMathOperator{\ketphi}{\ket|\phi>}
\DeclareMathOperator{\ketplus}{\ket|+>}
\DeclareMathOperator{\ketminus}{\ket|-->}
\DeclareMathOperator{\ketx}{\ket|x>}
\DeclareMathOperator{\tensor}{\otimes}

\DeclareMathOperator{\E}{\mathds{E}}

\DeclareMathOperator{\idm}{\mathds{1}}
\newcommand{\genemail}[2]{\href{#1}{#2}}

\begin{document}

\title{Generative AI Beyond Tokens:\\ Quantum Resource Consumption of IQP Circuits}

\author{\IEEEauthorblockN{Tom Krüger}
\IEEEauthorblockA{\textit{Technical University of Applied Sciences Regensburg}\\
\textit{FI CODE, Universität der Bundeswehr München}\\
Regensburg/Munich, Germany}
  \genemail{mailto:tom.krueger@othr.de}{tom.krueger@othr.de}
\and
\IEEEauthorblockN{Wolfgang Mauerer}
\IEEEauthorblockA{\textit{Technical University of Applied Science Regensburg}\\
        \textit{Siemens AG, Foundational Technology}\\
        Regensburg/Munich, Germany}
  \genemail{mailto:wolfgang.mauerer@othr.de}{wolfgang.mauerer@othr.de}
}

\maketitle

\begin{abstract}
Quantum generative modelling casts sampling as a generative task: a parametrised quantum circuit is trained such that sampling reproduces a target probability distribution. Instantaneous Quantum Polynomial-time (IQP) circuits combine structural simplicity with complexity-theoretic evidence for quantum advantage. Yet their practical value depends not only on expressivity, but on how efficiently they consume genuinely quantum resources.

We study this question through the lens of magic, or non-stabiliserness, as a resource for quantum generative modelling. We show that established fidelity- and geodesic-based notions of computational progress in  a projective Hilbert space are ill-suited to generative models, since operational performance is determined by output probability distributions rather than quantum states themselves. We evaluate magic-consumption directly on the probability simplex, using changes in Jensen-Shannon divergence to quantify progress. Applying this framework to trained random $\gamma$-sparse IQP circuits shows signatures of efficient magic use, with the dominant contribution arising from two-qubit gates. As IQP circuits produce remarkably low intermediate magic relative to phase-randomised states with the same sampling distributions, this renders IQP-based quantum generative models as promising candidates for resource-efficient demonstrations of quantum advantage on early fault-tolerant architectures.
\end{abstract}


\section{Introduction}
%Georgescu2022: Quantum Generative Adversarial Learning
%Quantum enhancement in generative models
% Protocols for classically training quantum  generative models on probability distributions

Quantum generative modelling (QGM) turns sampling quantum circuits into a generative task: a parameterised circuit is trained such that measurements of its output state reproduce a target probability distribution. This perspective is particularly natural in view of complexity-theoretic supremacy arguments that show certain quantum sampling protocols cannot be implemented efficiently classically~\cite{bouland2018quantum}. Instantaneous Quantum Polynomial-time (IQP) circuits are especially attractive: They provide restricted Born-machine ansätze for learning and generating probability distributions, while retaining the complexity-theoretic hardness of sampling from their output distributions under standard assumptions~\cite{Bremner2025,Bremner2017,pashayan2020estimation}.

Apart from the question if they can represent useful distributions, it is important to understand which genuinely quantum resources are required, particularly for fault-tolerant quantum computing (FTQC), where logical Clifford operations are comparatively benign, while  non-Clifford resources remain costly and must typically be supplied through magic-state preparation, injection, or distillation. Consumption of non-stabiliserness provides a means of determining   non-classical contributions to quantum generative modelling~\cite{Veitch2014,Leone2022,krueger:26:intermediateSRE}. Evaluating magic consumption allows for (a) physical cost estimation and (b) benchmarking the efficiency of converting quantumness into computational utility.

Existing approaches to magic-consumption for variational and structured circuits use geometric notions of progress in a projective Hilbert space~\cite{krueger:26:intermediateSRE}. For quantum generative models, however, distinct quantum states may induce the same measurement distribution, and are equivalent for generative performance; fidelity-based or geodesic measures on a Hilbert space can misrepresent progress in the learning task. We argue that magic-consumption evaluation must instead be formulated on the manifold of probability distributions, where the circuit output is compared to the target distribution. We realise this by replacing projective-state progress with a Jensen-Shannon divergence based measure on the probability simplex.

We analyse the efficiency of magic consumption in random $\gamma$-sparse IQP circuits trained as quantum generative models, and find that they exhibit signatures of efficient magic use; relevant contributions predominantly arise from two-qubit gates. As trained IQP circuits also produce intermediate states with remarkably low magic compared with phase-randomised states representing the same output distributions, we suggest that IQP-based quantum generative models offer a compelling architecture for benchmarking (and providing) resource-efficient quantum advantage on early FTQC devices.

\section{Preliminaries}
We fix notation and define basics in the following remarks.

\subsection{Quantum Generative Modeling}
Given a probability distribution $Q(x)$, 
quantum generative modelling~\cite{Rudolph2024,Barthe2025} seeks to train a parametrised quantum circuit $C(\phi)$ such that the output probability distribution $P(x) = \left|\braket<x|C(\phi)|0>\right|^2$ matches $Q(x)$. If $C(\phi)$ can be trained efficiently and $Q(x)$ is classically complex to sample from, QGMs can achieve quantum advantage.

\subsection{Sparse IQP}
\begin{definition}
A $n$ qubit IQP circuit is given by (a)
an input state $\ketzero$; (b) 
an implemented a unitary $\UIQP = H^{\otimes n} D H^{\otimes n}$, where $H^{\otimes n}$ is the $n$ qubit Hadamard gate and $D$ is a diagonal operator in the computational basis; and (c) a final state measurement in the computational basis.
\end{definition}

$D$ is commonly constructed from Pauli-$Z$ rotations $RZ_{I}(\theta)$ and controlled Pauli-$Z$ rotations $C_jRZ_I(\theta)$. Here $RZ_I(\theta) = \exp\ab(i \theta Z_I)$, where $Z_I$ is the Pauli string with $Z$ operators acting on all qubits $i \in I$; we also use   $Z_{I \cap [a:b]}$ for $Z_I$ restricted to qubits $a \leq i \leq b$. Further, $C_jRZ_I(\theta)$ is a $RZ_I(\theta)$ gate controlled by $j \notin I$. 

We rewrite $\UIQP$ by applying the Hadamard transformation to $D$: Uncontrolled $RZ_I(\theta)$ gates are transformed to $RX_I(\theta) = exp(i \theta X_I)$ gates, where $X_I$ is a tensor product of Pauli-$X$ operators acting on the qubits in $I \subset \mathds{N}$. For controlled $C_jRZ_I(\theta)$ gates, we assume qubits ${j} \cup I = \ab\{k_1, k_2, \dots, k_m\}$ are all adjacent and ordered such that $k_i < k_{i + 1}$ and $k_1 = j$ (if not, we permute wlog). Given such an order, $C_jRZ_I(\theta) = \diag(\idm, RZ_I(\theta))$. Consider a controlled $Z$ rotation $\diag(\idm, RZ(\theta) = \exp\ab(i \theta \diag(\mathbf{0}, Z))$ with $\diag(\mathbf{0}, Z) = (\idm Z - Z Z) / 2$. The Hadamard transform of the exponential of a diagonal operator $D$ is obtained by transforming $D$, thus $H \exp(i \theta \diag(\cdot)) H = \exp(i \theta H \diag(\cdot) H)$. In our case $H(\idm Z - Z Z) / 2 H = (\idm X - X X) / 2$ and $H C_jRZ(\theta) H = \exp\ab(i \theta (\idm X - X X) / 2)$.

Generalising to multi-qubits as $C_jRZ_I(\theta) = \exp\ab(i \theta \ab(Z_{I \cap [1:j-1]} \tensor \idm \tensor Z_{I \cap [j+1:n]} - Z_{I \cup \{j\}}) / 2)$. As gates in 
%the $Z$ rotation gate set 
\(\{RZ_I(\theta)\), \(C_jRZ_I(\theta)\}\) commute pairwise, a Hadamard transformed circuit constructed from this set can be expressed with gates out of \(\{RX_I(\theta)\), \(\exp(i \theta (X_{I \cap [1:j-1]} \tensor \idm \tensor X_{I \cap [j+1:n]} - X_{I \cup \{j\}}) / 2)\}\).

\textbf{Random $\gamma$-sparse IQP Circuits:} To sample a $\gamma$-sparse IQP circuit, for each pair of qubits one adds a 2-qubit gate with probability $\gamma \log(n) / n$, for $0 < \gamma \leq  n / \log(n)$ \cite{Paletta2024}. Additionally, a 1-qubit gate is added per qubit. On average, a random $\gamma$-sparse IQP circuit has $O(n \log(n))$ 2-qubit gates. In our case for a qubit pair $\{i, j\}$ we sampled the 2-qubit gates uniformly at random from $\{RZ_{\{i,j\}} (\theta), C_k RZ_l(\theta)\}$, where $(k,l) = (i,j)$ or $(j,i)$ with equal probability. Lastly, all gates where randomly shuffled to evenly distribute one-qubit and two-qubit gates.

\subsection{Magic}
We call the set of states reachable by Clifford operations the Clifford orbit or stabiliser states; they can be simulated efficiently classically. States beyond the Clifford orbit poses inherent non-stabiliser properties (non-stabiliserness) commonly referred to as \emph{magic}. Given the generally accepted hypothesis that non-stabiliser can not be simulated efficiently classically, magic serves as a resource theoretic measure \cite{Veitch2014} of the amount of actual quantum resources needed to prepare a state. Although, there are many proposed measures of magic, the $\alpha$-Stabiliser-\renyi-Entropy ($\sre_\alpha$) gained widespread adoption in recent years, for a detailed definition we refer to ~\cite{Leone2022}. 
Unless specified, we use $\alpha = 2$.

\subsection{Geometry}
The expected $\sre$ of a randomly $\Haar$ sampled $n$-qubit state $\E_{\ketpsi \sim \Haar} \ab(\sre_\alpha (\ketpsi)) \in O(n)$ for all $\alpha \geq 2$, which means that magic is a common feature of the average quantum state. Consequently, magic consumption of a quantum circuit is to be expected and does not necessarily contribute to the computational progress in a meaningful way. Therefore, the efficiency of magic consumption needs to be evaluated. In \cite{krueger:26:intermediateSRE} it was shown how based on the idea of geometric computational efficiency \cite{Anandan1990} the efficiency of magic consumption can be evaluated. Let $s_0\ab(\ketpsi, \ketphi)$ be the geodesic distance between $\ketpsi$ and $\ketphi$ on the projective Hilbert space.  Using $s_0$ as proposed in \cite{krueger:26:intermediateSRE} does not work in the case of QGM due to how quantum states are interpreted differently. To see this compare the two states $\ketplus$ and $\ketminus$, we have $s_0\ab(\ketplus, \ketminus) = \pi$. In QGMs states are interpreted as probability distributions where $P_{\ketpsi} (x) = \left|\braket<x|\psi>\right|^2$, clearly we have $P_{\ketplus} = P_{\ketminus}$. In this work we therefore propose to move the efficiency tracing of magic consumption to the space of probability distributions \ie the probability simplex. 

In information geometry, closeness between distributions is commonly evaluated with divergencies. A divergence does not fulfill all criteria of a metric, most importantly symmetry. Nevertheless, locally they coincide with metrics on the statistical manifold \cite{Amari2016}. A divergence widely adopted is the Kullback-Leibler divergence $\dkl (P : Q)$. Unfortunately $\dkl$ is not suited for our analysis as it is not well defined for $x$ s.t. $Q(x) = 0$ and $P(x) \neq 0$. Given two states $\ketpsi, \ketphi$, mapping them to the probability simplex, we can not guarantee this requirement. We propose using the Jensen-Shannon divergence 
\begin{equation}
    \jsd (P : Q) = \frac{1}{2} \ab(\dkl\ab(P : M) + \dkl\ab(Q : M))   
\end{equation}
with \(M = (P + Q)/2\), as $\jsd$ is well defined for all pairs of probability distributions mapped from quantum states, and is symmetric and bounded.

\subsection{Efficiency of Magic Consumption}
First we note that Magic is a state property, while resource consumption is a trait of state dynamics \eg circuits, gate operations, state propagation. A quantum gate $U$ only manipulates the magic of a state if $U$ is a non-Clifford operation. Vice versa, if $\sre(U\ketpsi) \neq \sre(\ketpsi)$, we know that $U$ performed a non-classical operation. Based on this observation, the \emph{magic consumption} of $U$ can be defined by $\ab|\Delta \sre| = \ab|\sre(U \ketpsi) - \sre(\ketpsi)|$ \cite{krueger:26:intermediateSRE}. The efficiency of said consumption can be evaluated by comparing it with the geometric approach towards the desired target. As we proposed above, in the case of QGMs a fitting measure is the delta in $\jsd$. In summary, the efficiency of magic consumption can be expressed by $\ab|\Delta \sre|$ vs $-\Delta \jsd$.

\section{Experiments / Simulations}
\label{sec:experiments}

To consider the efficiency of magic consumption in random $\gamma$-sparse IQP circuits, we generate random $\gamma$-sparse circuits and train them on random binomial mixture distributions (simulations are performed with QuantumToolbox~\cite{QuantumToolbox.jl2025}):
\begin{equation}
    \label{eq:mixBinDist}
    P(x ; p_1, \dots, p_4) = \frac{1}{4} \sum_{i=1}^4 P_{\textrm{Bin}}(x ; p_i)
    ,
\end{equation}
where $P_{\textrm{Bin}}(x; p_i)$ is the probability of the of $\sum_{k = 1}^n x_k 2^k \in \mathds{N}$ for $x \in \mathds{F}_2^n$ in a binomial distribution of $2^n -1$ draws with probability $p = p_i$.
We performed two batches of simulations: 1) to analyse the efficiency of magic consumption throughout the circuit and 2) to analyse the magic of intermediate states in the context of the degree of freedom allowed by relative qubit phases. 

\noindent\textbf{Batch 1:}
For the first batch we randomly generated 500 $7$-qubit circuit-distribution pairs $(C, P)$ for each $\gamma \in \{1, 1.4, 1.8, 2.2, 2.6, 3, 3.4\}$, here $C$ is a random $\gamma$-sparse IQP circuit as described above and $P$ is generated according to \cref{eq:mixBinDist} by sampling $0 < p_1, \dots, p_4 < 1$ uniformly at random. For each pair $C$ got optimised to produce a sampling distribution close to $P$. Concretely, the circuit was trained to minimise the $\jsd$ to $P$.

\noindent\textbf{Batch 2:}
A notable degree of freedom arises when mapping quantum states to probability distributions: Given a state $\ketpsi$ we obtain a distribution $P_{\ketpsi} (x) = \left|\braket<x|\psi>\right|^2$. Now consider a second state 
\begin{equation}
\label{eq:randPhState}
\ket|\psi(\mathbf{\theta})> = \sum_x e^{i \theta_x} \sqrt{P_{\ketpsi} (x)} \ketx ,
\end{equation}
then $P_{\ket|\psi(\theta)>} = P_{\ketpsi}$. To investigate this degree of freedom, we generated and trained 50 random circuit-distribution pairs $(C, P)$ similar to batch 1 with a fixed $\gamma = 1.5$. Let $C_k$ be the $k$-th gate of $C$, then $\ket|C_k> = C_k \cdots C_1 \ketzero$. For each $\ket|C_k>$ of the trained circuit we generated 100 phase-randomised states $\ket|C_k (\theta)>$ according to \cref{eq:randPhState} by choosing the $0 \leq \theta_k \leq 2\pi$ uniform at random.

\section{Results}
We again discuss results in two parts, as both batches capture different aspects of magic consumption efficiency.

\subsection{Batch 1}
In batch 1, we analyse magic consumption efficiency of each gate in $C$. Let again $\ket|C_k> = C_k \cdots C_1 \ketzero$ be the state after the $k$-th gate in $C$. We then calculate $\ab|\Delta_{k, k+1} \sre| = \left|\sre(\ket|C_{k+1}>) - \sre(\ket|C_k>)\right|$ and $\Delta_{k, k+1} \jsd = \jsd\ab(P_{\ket|C_{k+1}>}, P) - \jsd\ab(P_{\ket|C_k>}, P)$.
%All data points in the analysis of batch 1 were filtered for outliers defined by having a $z$-score $z(x) = (x - \mu) / \sigma$, with $\mu$ being the sample mean and $\sigma$ the standard deviation. A data point $x$ is classified as outlier if $|z(x)| > 3$. Furhter we filtred for $\left|\Delta \sre \right| > 0$ to ensure magic was consumed in that particular step.

\begin{figure}[htbp]
\includegraphics{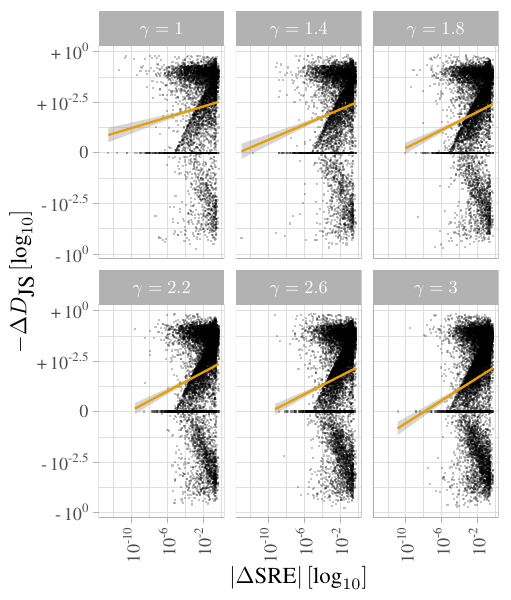}\vspace*{-1em}
    \caption{Magic consumption$\ab|\Delta \sre|$ is correlated with computational progress measured by $- \Delta \jsd$. It seems like a minimal density $1 < \gamma < 1.4$  is needed but for higher $\gamma$ values the correlation does not get stronger.}
    \label{fig:absdsreVsDJSperGamma}
\end{figure}

\begin{figure}[htbp]
\includegraphics{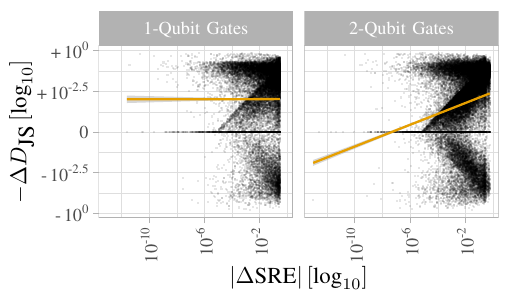}\vspace*{-1em}
    \caption{The majority of $1$-qubit gates contributes to computational progress measured in $- \Delta \jsd$, but without clear correlation to magic consumption $\ab|\Delta \sre|$. For $2$-qubit gates, a clear correlation pattern emerges.}
    \label{fig:absdsreVsDJSperGateType}
\end{figure}

\Cref{fig:absdsreVsDJSperGamma} shows that magic consumption $\left| \Delta \sre \right|$ correlates with  computational progress $-\Delta \jsd$. Further we note two jumps in magic efficiency at circuit densities $1 < \gamma < 1.4$ and $3 < \gamma < 3.4$.
\Cref{fig:absdsreVsDJSperGateType} shows a stark difference: The clear correlation between magic consumption and computational progress for 2-qubit gates is not present for 1-qubit gates.
A detailed split of correlation values over gate types and circuit densities is summarised in \cref{tab:correlations}. Additionally, the comparison to correlation values between magic consumption and geodesic distance changes $\Delta s_0$ on the projective Hilbert space further motivates our proposed evaluation metric $\Delta \jsd$.

\begin{table}
\sisetup{round-mode=places, round-precision=2}
\setlength{\tabcolsep}{2.5pt}
\centering
\caption{Correlation: Magic consumption and evaluation metrics.}
\begin{tabular}{lrrrrrrr}
    \toprule
    $\gamma$ & 1 & 1.4 & 1.8 & 2.2 & 2.6 & 3 & 3.4\\
    \midrule
    \multicolumn{8}{c}{All Gates} \\
    $\rho_{\ab|\Delta \sre|, -\Delta \jsd}$ &\SI{0.02866539}{} &\SI{0.06982176}{} &\SI{0.06988329}{} &\SI{0.07476883}{} &\SI{0.07351412}{} &\SI{0.06112293}{} &\SI{0.08888968}{}\\
    $\rho_{\ab|\Delta \sre|, -\Delta s_0}$  &\SI{0.008734128}{}  &\SI{0.021973906}{}  &\SI{0.021819540}{}  &\SI{0.005093901}{}  &\SI{0.011681930}{} &\SI{-0.003673601}{}  &\SI{0.018343343}{}\\
    \multicolumn{8}{c}{1-Qubit Gates} \\
    $\rho_{\ab|\Delta \sre|, -\Delta \jsd}$ &\SI{-0.059031642}{} &\SI{-0.009895621}{} &\SI{-0.034710928}{} &\SI{-0.037802382}{} &\SI{-0.027093384}{} &\SI{-0.034829768}{}  &\SI{0.005300185}{}\\  
    $\rho_{\ab|\Delta \sre|, -\Delta s_0}$  &\SI{0.016995401}{}  &\SI{0.033550209}{}  &\SI{0.022328450}{}  &\SI{0.010438683}{}  &\SI{0.007970203}{} &\SI{-0.020875559}{}  &\SI{0.035660938}{}\\
    \multicolumn{8}{c}{2-Qubit Gates} \\
    $\rho_{\ab|\Delta \sre|, -\Delta \jsd}$ &\SI{0.1096585}{} &\SI{0.1236942}{} &\SI{0.1295459}{} &\SI{0.1275139}{} &\SI{0.1090528}{} &\SI{0.0961843}{} &\SI{0.1152776}{}\\
    $\rho_{\ab|\Delta \sre|, -\Delta s_0}$&\SI{0.003703951}{} &\SI{0.017175875}{} &\SI{0.022910941}{} &\SI{0.002187109}{} &\SI{0.012814139}{} &\SI{0.002171129}{} &\SI{0.012157428}{}\\
    \bottomrule
\end{tabular}
\label{tab:correlations}
\end{table}

\subsection{Batch 2}
\begin{figure}[htbp]
\includegraphics{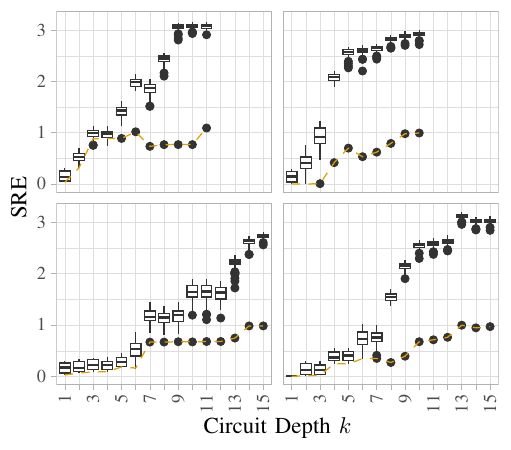}\vspace*{-1em}
    \caption{$\sre$ distribution of intermediate circuit states with random relative phases $\{\ket|C_k (\theta)>\}_\theta$. Dashed orange: $\sre$ of original in circuit states $\ket|C_k>$.}
    \label{fig:relPhaseSreDistr}
\end{figure}

\begin{figure}[htbp]
\includegraphics{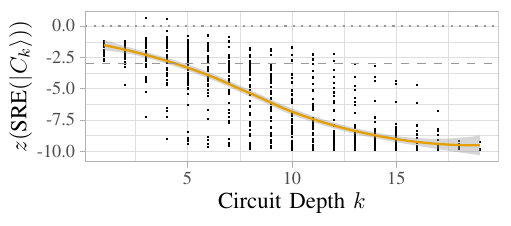}\vspace*{-1em}
    \caption{$z$-scores for the $\sre$ of each $\ket|C_k>$ in its batch of random $\{\ket|C_k (\theta)>\}_\theta$. All states $\ket|C_k>$ have a below average magic (below dotted line) and a majority of states can be categorised as statistical outliers (below dashed $z = -3$ line).}
    \label{fig:relPhaseSreZscore}
\end{figure}

Following up on the observation made above regarding the degree of freedom presented by alternative choices of relative phases, we will now investigate a second aspect of efficiency, when it comes to magic consumption. \Cref{fig:relPhaseSreDistr} shows the $\sre$ distribution of phase-randomised states $\ket|C_k(\theta)>$ for four selected circuits $C$ out of batch 2. States $\ket|C_k>$ produced by the trained circuit exhibit significantly less magic than the average state $\E_\theta(\ket|C_k(\theta)>)$. To further investigate this, we look at the $z$-score of $\sre(\ket|C_k>)$ within the distribution of $\{\sre(\ket|C_k(\theta)>)\}_\theta$. The $z$-score is defined as $z(x) = (x - \mu) / \sigma$, with $\mu$ being the sample mean and $\sigma$ the standard deviation. As shown in \cref{fig:relPhaseSreZscore}, all but two states $\ket|C_k>$ have a $z$-score $< 0$: Their magic is below average. The significant majority of states, especially at the later stages of the circuit, have a $z$-score $< -3$ and can be considered statistical outliers. 

\section{Conclusion}

We showed why geometric distance measures like $s_0$ on the projective Hilbert space are inapt to evaluate computational progress quantum generative models. Moreover, we propose to treat the quantum states as representations of probability distributions and to use $\jsd$ to replace $s_0$. Using this approach, we were able to evaluate the efficiency of magic consumption in random $\gamma$-sparse IQP circuits. Our analysis showed a correlation between magic consumption and computational progress in these circuits, with $2$-qubit gates seemingly being the driving factor of magic efficiency. Worth further investigation is the influence of the circuit density $\gamma$. The jumps in magic efficiency in the low and high $\gamma$ regimes sould be further investigated, to clarify whether they are mere experimental artifacts of low $\gamma$ resolution or if they alternatively hint at phase transitions in the magic efficiency of random $\gamma$-sparse IQP circuits. In our second batch of simulations we showed that random $\gamma$-sparse IQP circuits produce exceptionally low magic states compared to other states representing the same magic distribution. This further suggests IQP based quantum generative models to be a promising architecture for early fault tolerant systems.

\vspace{0.5em}

\noindent\textbf{Reproducibility}
A reproduction package~\cite{Mauerer:2022} can be found at \href{"https://github.com/lfd/qce26_IQP_magic_consumption"}{github.com/lfd/qce26\_IQP\_magic\_consumption} additionally a DOI safe version is available at \doi{10.5281/zenodo.21630858}.

\begin{small}
\noindent\textbf{Acknowledgment}
This work was supported by the German Federal Ministry of Research, Technology and Space (BMFTR), funding program \emph{quantum technologies---from basic research to market}, grant number 13N17387.
\end{small}

\printbibliography

\end{document}